\documentstyle[preprint,aps,prb,epsfig]{revtex}
\begin{document}

\title{Charge pairing by quantum entanglement in strongly correlated electron systems}

\author{Byung Gyu Chae}

\address{Basic Research Laboratory, Electronics and Telecommunications Research Institute,
Daejeon 305-700, South Korea}

\maketitle{}

\begin{abstract}

  Various charge pairings in strongly correlated electron systems are interpreted as quantum entanglement
of a composite system.
Particles in the intermediate phase have a tendency to form the coherent superposition state of
the localized state and the itinerant state,
which induces the entanglement of both particles in the bipartite subsystems
for increasing the entropy of the system.
The correction to the entropic Coulomb force becomes an immediate cause of charge pairing.

\end{abstract}

  The strong correlation in lattices brings out the dual behaviors of localization and itinerancy
of the charge carriers.\cite{1,2,3}
The coherent superposition state between the itinerant state and the localized state
can be generated by varying the correlation strength such as doping amount.
Analysis of complexity in the system has been attempted by considering the decoherence process of the coherent
state to the pointer states.\cite{4}
But, it has difficulty in explaining the charge pairing which leads to the superconducting phase
or the collective modes.

  In this study, we describe the possibility of forming the charge pairing by means of quantum entanglement
of a bipartite composite system.
The charge pairing is attributed to the entropy increase called as an amount of quantum entanglement.

  The mobile carriers in the correlated systems become localized in the lattice site
with increasing the correlation strength and finally it gets to the Mott insulator.
Most of the real system does not show an abrupt change in the transition as different
from the theoretical prediction,\cite{5,6}
and reveals the intermediate phase as like various collective modes.
We suggested in previous research\cite{4}
that the intermediate phase arises from the tendency to form the coherent superposition state
consisting of the ultimate basis states of the itinerant state, $|I$$>$ and the localized state, $|L$$>$.
Here, both states can be expanded as a linear combination of their basis vectors,
$\sum_{i}a_{i}|u_{i}$$>$ and $\sum_{j}b_{j}|v_{j}$$>$, respectively.

  We depict a coherent superposition state between the ultimate states in Fig. 1.
The coherent state $|\psi$$>_{C}$ of the particle is written as below,

\begin{equation} \label{eq:eps} |\psi$$>_{C} = \alpha |I$$> + \beta |L$$> \end{equation}

where $\alpha$ and $\beta$ can be represented as $\sqrt{Z}$ and $\sqrt{1-Z}$
by using the renormalization constant, $Z$.\cite{4}
In the real system, it is still obscure that
there is a concrete boundary between the coherent sate and both ultimate states.
However, we simply regard two particles in the coherent state $|\psi$$>_{C}$ as a bipartite composite system.
The total state of two particles is expressed in the tensor product form of the subsystems.

\begin{equation} \label{eq:eps} |\Psi$$> = |\psi$$>_{A} \otimes |\phi$$>_{B} \end{equation}

The density operator is a useful tool for analyzing this system.\cite{7,8}

\begin{equation} \label{eq:eps} \rho_{T} = |\Psi$$><$$\Psi| \end{equation}

where $\rho_{T}$ is the density operator of the composite system.
The reduced density operator of the subsystem $A$ can be obtained by taking the partial trace of $\rho_{T}$
over the basis of subsystem $B$.

\begin{equation} \label{eq:eps} \rho_{A} = \sum_{j} <$$j|_{B} (|\Psi$$><$$\Psi|) |j$$>_{B} = \rm{Tr}_{B} \rho_{T} \end{equation}

  In general,
the entangled pure state of the bipartite composite system leads to the mixed state of the subsystems.
That is, a bipartite pure state is entangled if and only if its reduced states are mixed.\cite{7}
We find that the entanglement accompanies the increase of the entropy of the total system.
This indicates that in reverse, the entropy increase may induce the entanglement of two particles.

Here, we consider one of the entangled states of the above composite system as follows,

\begin{equation} \label{eq:eps} |\Psi$$> = \frac{1}{\sqrt{2}}(|II$$> + |LL$$>) \end{equation}

where the $|I$$> \otimes |I$$>$ is abbreviated as $|II$$>$.
In this case, the total density operator is written as like,

\begin{equation} \label{eq:eps} \rho_{T} = \frac{1}{2}(|II$$><$$II| + |II$$><$$LL|
                                     + |LL$$><$$II| + |LL$$><$$LL|) \end{equation}

By taking the partial trace over the basis of subsystem $B$, we find that

\begin{equation} \label{eq:eps} \rho_{A} = \frac{1}{2}(|I$$><$$I| + |L$$><$$L|) \end{equation}

  This is the mixed state, which is a statistical ensemble of the itinerant state and the localized state.
Quantum entanglement makes the coherent superposition state decohere to the pointer states.
But, this does not indicate that the subsystem $A$ has a definite state of $|I$$>$ or $|L$$>$
before the external measuring process.

  As mentioned previously,
the coherent superposition state and the ultimate states can not clearly separate each other in the real system,
and thus coexist somewhat.
The ultimate states are comparatively stable because they exist in the correlation limitation.
Therefore these states play a role in a measuring device,
which invokes that the definite state of the particle is realized in the system.

  We extract that the entanglement of two particles brings out the charge pairing
because both particles on the subsystems are not only strongly correlated each other
but also the definite state.
The entangled states of the above composite system with two-level state may appear as
the form of complete Bell states.
The various types of charge pairing of electron-electron and electron-hole can be formed,
which may explain the complexity in strongly correlated electron systems.
Considering the entangled state in Eq. 5,
the particles in the itinerant states are entangled to the $|I$$>_{A}|I$$>_{B}$.
The entanglement of charge pair with an attractive force leads to the superconducting phase.

  The increase of the entropy is a cause of this entanglement
and becomes a measure of quantum entanglement.
The von Neumann entropy of the system is given by

\begin{equation} \label{eq:eps} S = - k_{B} \rm{Tr} (\rho \log \rho) \end{equation}

where $k_{B}$ is the Boltzmann constant.
Decoherence to the mixed states increase the entropy of the coherent subsystem,
which is the loss of information.
It is related that the system tends to progress in the direction of increasing entropy.
The increase of the entropy in this system is expressed by using the renormalization constant, $Z$.

\begin{equation} \label{eq:eps} S = - k_{B} [Z \log Z + (1-Z) \log (1-Z)] \end{equation}

  Figure 2 shows the change in the entropy according to the renormaliztion constant.
Here, the renormalization constant can be interpreted as the correlation strength
in strongly correlated electron system.
When the $Z$ is one half, the entropy, namely the amount of the entanglement has a maximum value.
In case the value of $Z$ is zero or one, there is no change in the entropy
because initially, the system does not have the intermediate phase
related to the coherent superposition state.

  In our description,
only the entropy and information can be interpreted as the origin of the charge pairing.
Recently,
it has been postulated that the entropic force is caused by
changes in the information associated with the positions of material bodies on a holographic screen.\cite{9}

\begin{equation} \label{eq:eps} F \Delta x = T \Delta S \end{equation}

where $T$ and $\Delta S$ are the temperature and the entropy change on the screen.
Gravity is explained as an entropic force on the basis of the holographic principle
and the equipartition rule of thermodynamics.
Naturally,
this scheme is expanded to other forces including Coulomb's law, the electroweak and strong forces.\cite{10,11}
Holographic principle representing the entropy-area relationship $ S = A/4l_{p}^2$ is a key concept.
Here, $A = 4 \pi R^2$ is the area of the spherical screen and $l_{p}^2$ is the Planck length.
However,
this entropy relationship is modified from the inclusion of quantum effects.

\begin{equation} \label{eq:eps} S = \frac{A}{4l_{p}^{2}} + s(A) \end{equation}

where $s(A)$ is the quantum correction term.
Power-law correction to the entropy was calculated by considering the entanglement of quantum fields.\cite{12}
The entropic Coulomb force is derived as below.\cite{13}

\begin{equation} \label{eq:eps} F = k \frac{Qq}{R^{2}} \left[1 + 4 l_{p}^{2} \frac{\partial s}{\partial A} \right] \end{equation}

  The first term is the usual Coulomb's law with Coulomb constant $k$ and charges $Q$, $q$ of the particles,
which is obtained from the entropy-area relationship.
The second term is the corrected entropic Coulomb force by entanglement of quantum fields
in a superposition state.
Here,
even though the further study is required,
the pairing force of our model is the corrected entropic Coulomb force
because it is resulted from the entanglement of the coherent superposition states.
If the usual Coulomb force is sufficiently screened in the real system,
the corrected entropic force may be enough for the charges to be coupled.

  Now,
the entropic analysis of the fundamental force is progressive subject.
It is not easy to find the physical system applying for this formalism.
According to above description of the charge pairing in the intermediate phase,
our model can be a typical system where the correction to entropic Coulomb force is realized.

  In conclusion,
we find that
the intermediate phase in strongly correlated electron system
is attributed to the tendency to form the coherent superposition state.
For increasing the entropy of the system,
two particles are coupled by means of quantum entanglement.
The complexity such as superconducting pairing or charge stripe can be understood
by this scheme.
Charge pairing is resulted from the entropic correction to Coulomb's law.

\begin{figure}
\vspace{2.0cm}
\centerline{\epsfysize=7cm\epsfxsize=9cm\epsfbox{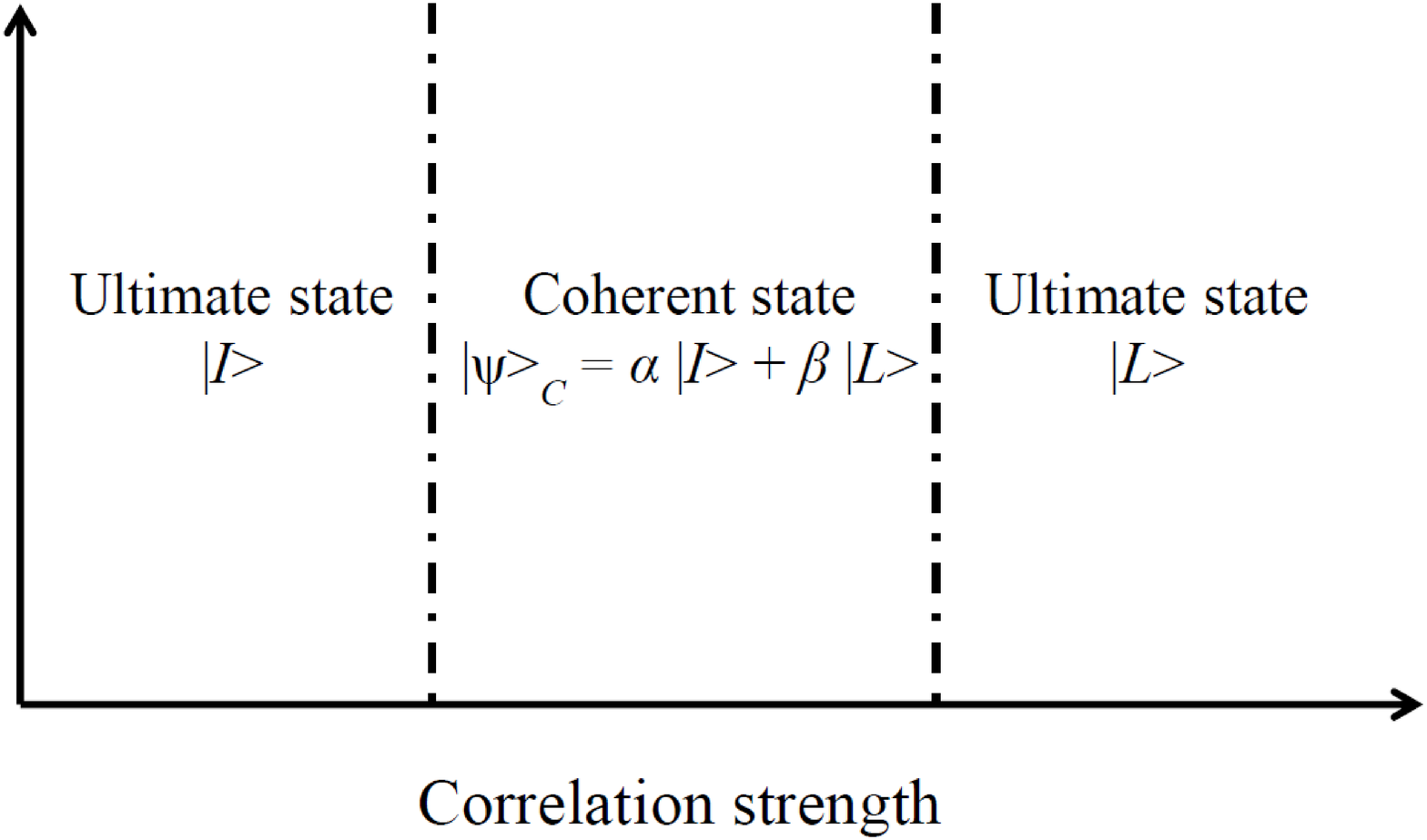}}
\vspace{0.0cm}
\caption{A schematic diagram for a coherent superposition state between the ultimate states.}
\label{f1}
\end{figure}

\begin{figure}
\vspace{2.0cm}
\centerline{\epsfysize=7cm\epsfxsize=10cm\epsfbox{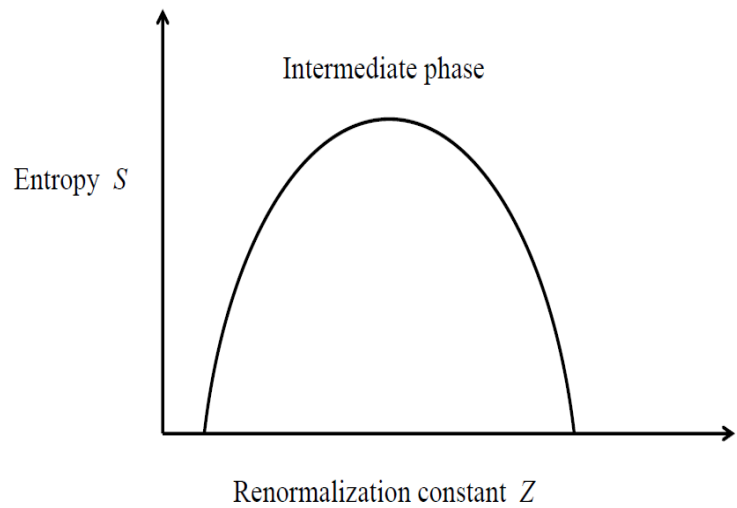}}
\vspace{0.0cm}
\caption{Change in the entropy according to the renormaliztion constant.}
\label{f1}
\end{figure}

\end{document}